\documentstyle[amssymb,preprint, aps]{revtex}
\begin{document}
\draft
\title{Chaotic atomic population oscillations between two coupled Bose-Einstein
condensates with time-dependent asymmetric trap potential }
\author{Chaohong Lee, Lei Shi, Xiwen Zhu, Kelin Gao}
\address{\ State Key Laboratory of Magnetic Resonance and Atomic and Molecular
Physics,\\
Wuhan Institute of Physics and Mathematics, The Chinese Academy of Sciences,%
\\
Wuhan, 430071, P. R. China }
\author{Wenhua Hai, Yiwu Duan}
\address{Department of Physics, Hunan Normal University, Changsha, 410081, P. R. China}
\author{Wing-Ki Liu}
\address{Department of Physics, University of Waterloo, Waterloo, Ontario,
N2L3G1,Canada}
\date{\today}
\maketitle

\begin{abstract}
We have investigated the chaotic atomic population oscillations between two
coupled Bose-Einstein condensates (BEC) with time-dependent asymmetric trap
potential. In the perturbative regime, the population oscillations can be
described by the Duffing equation, and the chaotic oscillations near the
separatrix solution are analyzed. The sufficient-necessary conditions for
stable oscillations depend on the physical parameters and initial conditions
sensitively. The first-order necessary condition indicates that the Melnikov
function is equal to zero, so the stable oscillations are Melnikov chaotic.
For the ordinary parameters and initial conditions, the chaotic dynamics is
simulated with numerical calculation. If the damping is absent, with the
increasing of the trap asymmetry, the regular oscillations become chaotic
gradually, the corresponding stroboscopic Poincare sections (SPS) vary from
a single island to more islands, and then the chaotic sea. For the
completely chaotic oscillations, the long-term localization disappears and
the short-term localization can be changed from one of the BECs to the other
through the route of Rabi oscillation. When there exists damping, the
stationary chaos disappears, the transient chaos is a common phenomenon
before regular stable frequency locked oscillations. And proper damping can
keep localization long-lived.
\end{abstract}



\section{Introduction}

The study of Bose-Einstein condensate (BEC) in gases can give new
understanding of atomic, condensed-matter, and statistical physics. Due to
the remarkable development in the ability to control atomic motion by
optical techniques, BEC was first produced in a weakly interacting gas of
alkali metal atoms held in magnetic trap in 1995 [1]. Following the first
observation, many important experiments were carried out. The group of
Durfee and Ketterle detected the condensate in cooled gases, the collective
excitations, the collisions between separately-prepared condensates, and the
pulsed output of a prototype atom laser [2,3], they also observed the
interference between two Bose condensates and demonstrated that Bose
condensed atoms are $``$laser-like$"$, that is, they are coherent and show
long-range correlations [4]. These results have direct implications for the
atom laser and the Josephson effect for atoms. Recently, solitons have been
generated in BEC by properly phase imprinting [5], the phase of a BEC
wave-function was prepared with optical imprinting techniques and measured
with a Mach-Zehnder matter-wave interferometer that makes use of optically
induced Bragg diffraction.

Experimental achievements of BEC caused great theoretical interests in this
novel field. There exists abundant nonlinear dynamics in BEC, for the
macroscopic condensate wave function obeys a nonlinear equation, which is
called as the Gross-Pitaveskii equation (GPE) [6]. Vortex stability in BEC
is explained by using a two-mode model [7]. Oscillations of atomic
populations and collective excitations at high energies was detailed [8],
the complex dependencies of these excitation energies bring us close to the
notion of chaos, and the role of chaotic motions in the dynamics of BEC
remains to be studied. Large amplitude oscillations of condensed trapped
atoms to external driving magnetic fields was analyzed by the group of
Smerzi [9], their results of frequencies and excitation times of collective
oscillations consistent with the experimental data very much. An appropriate
semiclassical limit for GPE with an additional chaotic potential was given
out by using a semiclassical interpretation of the Wigner function [10].
Assuming the background density and velocity vary slowly on the soliton
scale, Busch derived the equation of motion for dark soliton propagating
through an effectively one-dimensional cloud of BEC, by using a multiple
scale boundary layer theory. Recently, Filho et al. investigated the
dynamics of the growing and collapsing of BEC in a system of trapped
ultracold atoms with negative scattering lengths and found that the number
of atoms can go far beyond the static stability limit [12].

Current efforts are being focused on coupled two-component and
multi-component BEC. Interference and dynamics of component separation in
two-component BEC was observed [4,13]. The quantum statistics of the ground
state of a two-mode model for coupled BECs was analyzed [14], and strong
squeezing of the number difference for positive nonlinearities and a regime
of squeezing in the relative phase for negative nonlinearities were
revealed. The dynamics of Josephson-like oscillations between two coupled
BECs was studied using the time-dependent variational method [15], from the
calculation result, the tunneling dynamics is $``$coherent$"$ when the trap
is not displaced, that is, the orbitals of each condensate do not change; on
the other hand, the change in the condensate orbitals has a strong influence
in the tunneling dynamics when the trap is displaced. Smerzi and Raghavan
researched the coherent atomic tunneling and oscillations between two
zero-temperature BECs confined in a double-well magnetic trap in the case of
weakly Josephson coupling [16,17]. The coupling was provided by a laser
barrier in a double well magnetic trap or by Raman coupling between two
condensates in different hyperfine levels. The dynamics of phase difference
and fractional population imbalance was described with the two-mode
nonlinear GPE called as Bose Josephson junction (BJJ) equation. In addition
to the nonsinusoidal anharmonic generalization of the ac Josephson effect
and plasma oscillations occurring in the superconductor junction (SJJ), the
macroscopic quantum self-trapping (MQST, a self-maintained population
imbalance with nonzero average value of the fractional population
imbalance.) and the $\pi -$phase oscillations ( the time averaged value of
the phase difference is equal to $\pi $) were also observed.

In the case of time-dependent trapping potential and non-negative damping
and finite temperature effects, the more interesting nonlinear dynamics
emerges out, such as chaos. Abdullaev and Kraenkel analyzed the coherent
atomic oscillations and resonances between two coupled BECs in a double-well
trap with time-dependent tunneling amplitude for different damping [18].
With a slowly varying trap, the nonlinear resonances and chaos exist in the
oscillations of the fractional imbalance. The conditions for chaotic
macroscopic quantum tunneling phenomena were obtained with the use of the
Melnikov function approach, and the chaotic oscillations depend on the
frequency and modulation amplitude sensitively. For the rapidly varying
case, the averaged system was given out by using the multiscale expansion
method. They also considered the macroscopic quantum tunneling and
resonances in coupled BECs with oscillating atomic scattering length [19].
The chaotic oscillations in the relative atomic population due to the
overlaps between nonlinear resonances were showed. And the possibility of
stabilization of the unstable -mode regime was derived from the analyzing of
the oscillations in the rapidly varying case.

We know the laser barrier position and the laser beam intensity of the laser
beam in the trap can be modified in experiments, so the trap asymmetry and
the amplitude of the tunneling between the coupled BECs can be
time-dependent. In our present paper, we will analyze the chaotic
oscillation of the fractional population imbalance between two Josephson
coupled BECs with time-dependent asymmetric trap potential, using both
analytical and numerical approach. The structure of this paper is as
follows. In this section, we briefly review both experimental and
theoretical developing of BEC, and also show the purpose of the paper. The
effective particle model for the oscillations of the fractional population
imbalance is derived from the two-mode GPE in the next section. In the case
of small time-dependent trap asymmetry and small damping, using the
analytical method [20,21] developed by us, the chaotic dynamics of the
fractional imbalance near the separatrix solution is analyzed in details in
the third section. The conditions for chaotic oscillation and criteria for
the onset of chaos are also obtained. And the regions of regular and chaotic
oscillation are showed. In the fourth section, the population oscillations
are simulated with numerical approach for arbitrary time-dependent trap
asymmetry and damping. In the last section, a brief precise summary and
discussion are given out.

\section{The effective particle model for the oscillations of the fractional
population imbalance}

Ignoring the damping and finite-temperature effects, the problem of coupled
Bose-Einstein condensates in a double well trap can be described with the
following nonlinear two-mode dynamical equations 
\begin{eqnarray}
i\hbar \frac{\partial \psi _{_{1}}}{\partial t} &=&[E_{1}+U_{1}\left| \psi
_{_{1}}\right| ^{2}]\psi _{_{1}}-K\psi _{_{2}},  \nonumber \\
i\hbar \frac{\partial \psi _{_{2}}}{\partial t} &=&[E_{2}+U_{2}\left| \psi
_{_{2}}\right| ^{2}]\psi _{_{2}}-K\psi _{_{1}}.
\end{eqnarray}
Where, $E_{1}$ and $E_{2}$ are zero-point energies in each well, $%
U_{1}\left| \psi _{_{1}}\right| ^{2}$ and $U_{2}\left| \psi _{_{2}}\right|
^{2}$ are proportional to the atomic self-interacting energies, and $K$
describes the amplitude of the tunneling between two condensates. These
parameters are defined by the overlap integrals of the time-dependent
Gross-Pitaevsky eigenfunctions. The above equations are named as BJJ
equations, which have been derived from GPE in reference [17].

The wave function $\psi _{i}$ can be written in the form of $\psi _{_{i}}=%
\sqrt{N_{i}}\exp (i\theta _{i})$, here, $N_{i\text{ }}$and $\theta _{i}$ are
the amplitudes for general occupations and phases respectively. Then the
fractional population imbalance can be defined as 
\begin{equation}
z(t)=\frac{N_{1}(t)-N_{2}(t)}{N_{T}}=\frac{\left| \psi _{_{1}}\right|
^{2}-\left| \psi _{_{2}}\right| ^{2}}{\left| \psi _{_{1}}\right| ^{2}+\left|
\psi _{_{2}}\right| ^{2}}.
\end{equation}
here, $N_{T}=N_{1}(t)+N_{2}(t)$ is a constant total number of condensed
atoms. And the relative phase is 
\begin{equation}
\phi (t)=\theta _{2}(t)-\theta _{1}(t).
\end{equation}
Then the fractional population imbalance and the relative phase obey the
following differential equations 
\begin{eqnarray}
\frac{dz}{dt} &=&-\frac{2K}{\hbar }\sqrt{1-z^{2}}\sin \phi ,  \nonumber \\
\frac{d\phi }{dt} &=&\frac{2K}{\hbar }(\Delta E+\Lambda z+\frac{z}{\sqrt{%
1-z^{2}}}\cos \phi ).
\end{eqnarray}
The parameters $\Delta E$ and $\Lambda $ determine the dynamic regimes of
the BEC atomic tunneling and they can be expressed as 
\begin{equation}
\Delta E=\frac{E_{1}-E_{2}}{2K}+\frac{U_{1}-U_{2}}{4K}N_{T}.
\end{equation}
\begin{equation}
\Lambda =UN_{T}/2K 
\begin{array}{lll}
& , & 
\end{array}
U=(U_{1}+U_{2})/2.
\end{equation}
Because of the overlapping condensate, there exists different kind of
damping for different type of overlapping. For examples, if we take into
account a noncoherent dissipative current of normal-state atoms, the
differential equation which describes the oscillations of the atomic
population including the damping term $-\eta d\phi /dt$; and the damping has
the form $-\eta z(t)$ for the two interacting condensates with different
hyperfine levels in a single harmonic trap. In the case of time-independent
parameter $K$, we can rescale $(\frac{2K}{\hbar })t$ to a dimensionless time 
$t$. The motion of the fractional population imbalance and relative phase is
very similar to a nonrigid pendulum. The Hamiltonian of the unperturbed $%
(\Delta E=cons\tan t,\eta =0)$ dimensionless system is as follows 
\begin{equation}
H=\frac{\Lambda z^{2}}{2}+\Delta Ez-\sqrt{1-z^{2}}\cos \phi .
\end{equation}
The corresponding canonical equations of the motion are equivalent to the
equations of the oscillations of the atomic population imbalance and
relative phase, their forms are 
\begin{equation}
\frac{d\phi }{dt}=\frac{\partial H}{\partial z} 
\begin{array}{lll}
& , & 
\end{array}
\frac{dz}{dt}=-\frac{\partial H}{\partial \phi }.
\end{equation}
For the time-independent trapping potential, the energy of the above system
is conservative.

In order to see the oscillations of the fractional population imbalance more
transparently, we introduce an effective classical particle whose coordinate
is $z$, moving in a potential $V$ with the initial energy $E_{eff}^{0}=[%
\stackrel{.}{z}(t)|_{t=t_{0}}]^{2}/2+V|_{t=t_{0}}$. If the trapping
potential is time-independent, the effective potential $V$ is
time-independent too, and the motion of the effective particle is regular;
otherwise, for the time-dependent trapping potential, $V$ varies with time
and the corresponding motion is anharmonic and even chaotic. The chaotic
dynamics in atomic tunneling will be detailed in the following sections. For
the time-independent constant coupling $K$ , the Hamiltonian of the
effective particle is given by 
\begin{eqnarray}
H_{eff} &=&\frac{1}{2}p_{z}^{2}+V=\frac{1}{2}(1-H^{2})  \nonumber \\
V &=&\frac{1}{2}z^{2}(1-\Lambda H+\frac{1}{4}\Lambda ^{2}z^{2})+(\frac{1}{2}%
\Delta E\Lambda z^{3}+\frac{1}{2}\Delta E^{2}z^{2}-\Delta EHz).
\end{eqnarray}
The Hamiltonian's canonical equations of the motion are 
\begin{equation}
\frac{dz}{dt}=\frac{\partial H_{eff}}{\partial p_{z}} 
\begin{array}{lll}
& , & 
\end{array}
\frac{dp_{z}}{dt}=-\frac{\partial H_{eff}}{\partial z}.
\end{equation}
For the symmetric trapping potential $(\Delta E=0)$, the effective potential 
$V$ is time-independent, increasing the value of $(1-\Lambda H)$ from
negative to positive changes the effective potential from a double-well to a
parabolic. The effective particle moves between the classical turning
points, where the kinetic energy of the effective particle is zero. Fig. 1
shows the changing of the shape of the effective potential, $(A)$ for
different values of $H$ with fixed value of $\Lambda $, and $(B)$ for
different values of $\Lambda $ with fixed value of $H$.

The motion in the parabolic potential is Rabi oscillation with a zero
time-average value of the fractional population imbalance $z$. For fixed
parameters $\Lambda $ and $H$, the oscillations with small effective
energies $H_{eff\text{ }}$are sinusoidal, the increasing of the effective
energies adds higher harmonics to the sinusoidal oscillations.

In the case of double-well potential, the motion is very different from the
case of the parabolic potential. If the effective energies is greater than
the barrier between two wells, that is, $H_{eff}>0$, the motion is a
nonlinear Rabi oscillation with a zero time-average value of $z$, it
corresponds to the periodic flux of atoms from one BEC to the other. If the
effective energy is little than the potential barrier, $H_{eff}<0$, the
particle is confined in one of the two wells, it means the localization of
atomic population in one of the two condensates, and this localizing
phenomenon has been named as macroscopic quantum self-trapping (MQST). At
the threshold point, the effective energy is equal to the potential barrier, 
$H_{eff}=0$, the corresponds threshold motion separating the above two
regimes, the separatrix solution for the right-hand side well is 
\begin{equation}
z_{s}(t)=2\sqrt{(\Lambda H-1)/\Lambda ^{2}}\sec h\xi 
\begin{array}{lll}
& , & 
\end{array}
\xi =C_{0}+t\sqrt{\Lambda H-1}.
\end{equation}
Where, constant $C_{0}$ is determined by the initial conditions. Considering
the physical constrain, the amplitude of the fractional atomic population
imbalance oscillations must be little than one, i.e.,$\left| z\right| _{\max
}\leqslant 1$ , so the abstract value $z$ of at points with the lowest
potential energy must be little than one too, this requires the parameters
satisfy $\sqrt{2(\Lambda H-1)/\Lambda ^{2}}\leqslant 1$, the atoms
completely localize on one of the two condensates when $\sqrt{2(\Lambda
H-1)/\Lambda ^{2}}=1$. And if the amplitude of the separatrix solution is
lager than one, i.e., $2\sqrt{(\Lambda H-1)/\Lambda ^{2}}>1$, then there
only exists MQST.

Based upon the above analysis, we know that the pitchfork bifurcation occurs
at the point with $\Lambda H=1$ for the time-independent symmetric trapping
potential, that is to say the equilibrium point at the origin changes
stability type and two new additional equilibrium points are created. For
the asymmetric and time-dependent trapping potential, the bifurcation
becomes more complex, the regular oscillations become chaotic through the
route of period doubling.

\section{Chaotic oscillations near the separatrix solution with small trap
asymmetry}

It is very interesting to investigate the dynamics of the fractional
population imbalance near the separatrix solution, that is, the initial
conditions and physical parameters are very close to the separatrix of the
unperturbed symmetric system. In the case of small trap asymmetry $\Delta E$
and small damping $\eta $, they can be looked as perturbations to the
symmetric system, from the general theory of nonlinear driven oscillations,
the chaotic macroscopic quantum tunneling phenomena appears when the trap
asymmetry is time-dependent. For the two interacting condensates with
different hyperfine levels in a single harmonic trap, the damping has the
form $-\eta dz/dt$, driving from the effective potential in the previous
section, the Newtonian equation of the motion for the fractional population
imbalance is given as the following Duffing equation. 
\begin{equation}
\frac{d^{2}z}{dt^{2}}-(\Lambda H-1)z+\frac{\Lambda ^{2}}{2}z^{3}=-\frac{3}{2}%
\Delta E\Lambda z^{2}-\Delta E^{2}z+\Delta EH-{\normalsize \eta }\frac{dz}{dt%
}.
\end{equation}
In addition to a time-independent trap asymmetry $\Delta E_{0}$, we impose a
sinusoidal variation so that we can write the asymmetry term as $\Delta
E=\Delta E_{0}+\Delta E_{1}\sin \omega t$. When the intensity of the laser
beam is fixed, varying the laser barrier position can realize this. Writing
the trap asmmetry and the damping as following form 
\begin{equation}
\Delta E=\Delta E_{0}+\Delta E_{1}\sin \omega t=\varepsilon (F_{0}+F_{1}\sin
\omega t) 
\begin{array}{lll}
& , & 
\end{array}
\eta =\varepsilon \mu .
\end{equation}
In the above, $\varepsilon $ is a dimensionless parameter. Using the
analytical approach developed by us [20,21], we write the solution close to
the separatrix solution as the following expansions 
\begin{equation}
z=\sum\limits_{i=0}^{+\infty }\varepsilon ^{i}z_{^{i}}=z_{_{0}}+\varepsilon
z_{_{1}}+\varepsilon ^{2}z_{_{2}}+...
\end{equation}
Here, $z_{^{i}}$ are the $i-th$ order corrections. Substituting the above
expression into the Newtonian equation of the motion, comparing the
coefficient function of every $\varepsilon ^{i}$ of both sides of the
differential equation, setting $\varepsilon $ as $1$, then we obtain $%
z_{^{i}}$ satisfy 
\begin{equation}
\frac{d^{2}z_{0}}{dt^{2}}-(\Lambda H-1)z_{0}+\frac{\Lambda ^{2}}{2}%
z_{0}^{3}=0.
\end{equation}
\[
\frac{d^{2}z_{i}}{dt^{2}}-(\Lambda H-1)z_{i}+\frac{3\Lambda ^{2}}{2}%
z_{0}^{2}z_{i}=\epsilon _{i}, 
\]
\begin{equation}
\epsilon _{_{1}}=-\eta \frac{dz_{_{0}}}{dt}+\Delta EH-\frac{3}{2}\Delta
E\Lambda z_{0}^{2},\epsilon _{_{2}}=-\eta \frac{dz_{_{1}}}{dt}-3\Delta
E\Lambda z_{_{0}}z_{_{1}}-\frac{3}{2}\Lambda
^{^{2}}z_{_{0}}z_{_{1}}^{^{2}},...
\end{equation}
The zero-order solution is the separatrix solution, and the basic solutions
of the high-order corrections are as follows 
\begin{equation}
z_{i1}^{0}=\frac{dz_{0}}{dt}=-\frac{2(\Lambda H-1)}{\sqrt{\Lambda ^{2}}}\sec
h\xi \tanh \xi .
\end{equation}
\begin{eqnarray}
z_{i2}^{0} &=&z_{i1}^{0}\int (z_{i1}^{0})^{-2}dt  \nonumber \\
&=&\frac{-\sqrt{\Lambda ^{2}}}{16(\Lambda H-1)^{3/2}}\sec h^{2}\xi (\cosh
3\xi -9\cosh \xi +12\xi \sinh \xi ).
\end{eqnarray}
So the general expressions of $i-th$ corrections are in form of 
\begin{equation}
z_{i}=z_{i2}^{0}\int\limits_{C_{1}}^{t}z_{i1}^{0}\epsilon
_{i}dt-z_{i1}^{0}\int\limits_{C_{2}}^{t}z_{i2}^{0}\epsilon _{i}dt.
\end{equation}
Constants $C_{1}$ and $C_{2}$ are determined by the initial conditions and
physical parameters. Apparently, $\left| z_{i1}^{0}\right| \rightarrow 0$
and $\left| z_{i2}^{0}\right| \rightarrow +\infty $, when time $t\rightarrow
\pm \infty $. Solving the $i-th$ order equations one by one, we can obtain $%
\epsilon _{i}$ are time-periodic functions with finite amplitudes. This
means the high-order corrections are non-convergent unless the coefficient
functions of the growing function $z_{i2}^{0}$ are equal to zero. So the
general motion is unstable periodic oscillations, the necessary-sufficient
conditions for stable oscillations are 
\begin{equation}
\lim_{t\rightarrow \pm \infty }\int\limits_{C_{1}}^{t}z_{i1}^{0}\epsilon
_{i}dt=0.
\end{equation}
The above conditions are non-integrable, clearly, they contain the following
necessary conditions 
\begin{equation}
\int\limits_{-\infty }^{+\infty }z_{i1}^{0}\epsilon _{i}dt=0.
\end{equation}
Apparently, the first integral $(i=1)$ of the necessary conditions is the
Melnikov function of the system, the necessary conditions indicate that the
Melnikov function is equal to zero; this means that the stable oscillations
are Melnikov chaotic. But because of the non-sufficient property of the
above condition, not all chaotic oscillations are stable. Integrating the
above equations, one can obtain the necessary conditions are a series of
relations of the initial conditions and parameters, for fixed initial
conditions, modifying the parameters can control the instability of the
chaotic oscillations. Substituting the expressions of $z_{11}^{0}$ and $%
\epsilon _{1}$ into the necessary condition, integrating it yields the
first-order condition 
\[
-\frac{8\eta (\Lambda H-1)\sqrt{\Lambda H-1}}{3\Lambda ^{2}}-\frac{2\Delta
E_{1}H}{\sqrt{\Lambda ^{2}}}\omega \pi \cos \frac{\omega C_{0}}{\sqrt{%
\Lambda H-1}}\sec h\frac{\omega \pi }{2\sqrt{\Lambda H-1}} 
\]
\begin{equation}
+\frac{2\Delta E_{1}(\Lambda H-1)\sqrt{\Lambda H-1}}{\Lambda \sqrt{\Lambda
^{2}}}(1+\frac{\omega ^{2}}{\Lambda H-1})\omega \pi \cos \frac{\omega C_{0}}{%
\sqrt{\Lambda H-1}}\sec h\frac{\omega \pi }{2\sqrt{\Lambda H-1}}=0.
\end{equation}
The above necessary condition is irrelative to the time-independent trap
asymmetry $\Delta E_{0}$, this means the chaotic oscillations are caused by
the time-dependent trap asymmetry, but it is not to say that the stability
is irrelative to the time-independent trap asymmetry, actually, the
sufficient-necessary conditions and high-order necessary conditions are
relations of it and other parameters. For the same parameters, the
distribution of stability curves sensitive depends on the initial
conditions, to show explicitly this dependence we have chosen a series of
value of the initial constant $C_{0}$, with the growing of the value of $%
C_{0}$, the curves become denser and denser, this illustrates the existence
of chaos, see Fig. 2. The changing between the regular oscillations and the
chaotic oscillations is showed in Fig.3. Regions above the curves correspond
to chaotic oscillations of the fractional population imbalance and those
below correspond to regular oscillations. There exist two chaotic regions
separated by a special frequency which is determined by the physical
parameters, and this frequency can cause an unstable nonlinear resonance.
When the damping becomes stronger and stronger, the regions of chaotic
oscillations become smaller and smaller, and the regular region becomes
larger and larger.

\section{Numerical simulation}

In general, the atomic population oscillation is far away from the
separatrix solution, and then the oscillation dynamics can not be obtained
from the previous analytical method. In this section, using the fourth
Runge-Kutta method with variable step-width, the chaotic population
oscillations are simulated by straightforward numerical integration of the
motion equations of the dimensionless model of system (4) with constant
parameter $K$, and the trap asymmetry is in form of $\Delta E=\Delta
E_{0}+\Delta E_{1}\sin \omega t$, and the damping of the population
oscillation is $-\eta z(t)$. In the time -independent symmetric trap $%
(\Delta E=0)$, because of the damping, both Rabi oscillation and MQST reach
an equilibrium state with zero population imbalance, see Fig. 4 (B) and (F),
for the time-independent asymmetric case, the equilibrium state departure
from the zero population imbalance, see (D) and (H) of Fig.4. Ignoring the
damping effects, the oscillations are regular, they contain two different
kinds, Rabi oscillation and MQST,see (A), (E), (C) and (G) of Fig.4.

For the time-dependent asymmetric trap potential, the chaotic oscillation
emerges out. Sampling a single trajectory every period of the varying of the
trap asymmetry, then we can obtain the stroboscopic Poincare section (SPS).
When the damping is absent, with the increasing of the time-dependent trap
asymmetry $\Delta E_{1}$, the sections vary from a single island into a lot
of islands, and at last all islands are submerged by the chaotic sea. This
means the periodic oscillations become quasi-periodic, and then chaotic.
Fig. 5 is the SPS of $(z,dz/dt)$, with $z(0)=0.5$, $\phi (0)=0.0$, $\Delta
E_{0}=0.0$, $\Lambda =10.0$, $\omega =4\pi $ and $\eta =0.0$, for these
initial conditions if the damping and the trap asymmetry are absent, the
corresponding oscillation is Rabi oscillation. When $\Delta E_{1}=3.000,$
there is only a single island. Then it is separated into six islands when $%
\Delta E_{1}$ increase to $6.000$. For larger trap asymmetry, $\Delta
E_{1}=6.750$, the regular islands are surrounded by the chaotic sea. For
large enough trap asymmetry, $\Delta E_{1}=7.500$, the regular islands are
all submerged by the chaotic sea, and the sea is symmetrical to $z=0$.
Starting from the MQST, the SPS with $z(0)=0.75$, $\phi (0)=0.0$, $\Delta
E_{0}=0.0$, $\Lambda =10.0$, $\omega =2\pi $ and $\eta =0.0$ is showed in
Fig. 6, the similar dynamics is exhibited. For small $\Delta E_{1}$ $%
(1.000,1.560$ and $1.565)$, the time-averaged value of the fractional
population imbalance is non-zero, the atoms are localized on one of the
condensates. However, for large enough $\Delta E_{1}$ $(1.700)$, the chaotic
sea is symmetrical to $z=0$. This indicates that, in the completely chaotic
oscillation, the time-averaged value of the fractional population imbalance
is zero, and the long-lived MQST or localization disappears.

The completely chaotic oscillations of the fractional population imbalance
from Rabi oscillation and MQST are presented in Fig. 7. The left column
corresponds to $z(0)=0.5$, $\phi (0)=0.0$, $\Delta E_{0}=0.0$, $\Lambda
=10.0 $, $\omega =4\pi $, $\Delta E_{1}=7.500$ and $\eta =0.0$, the right
column corresponds to $z(0)=0.75$, $\phi (0)=0.0$, $\Delta E_{0}=0.0$, $%
\Lambda =10.0$, $\omega =2\pi $, $\Delta E_{1}=1.700$ and $\eta =0.0$. The
first row is the time evolution of $z$, the second row is the power spectra
of the corresponding oscillation. Apparently, through the tunnel of Rabi
oscillation, the short-term localization or MQST can be changed from one of
the BECs to the other, and the corresponding power spectra is continuous.

Because of the existence of the damping, the dimensionless system is not a
Hamiltonian system but a dissipative system and the volume in phase space
will decrease through time evolution. Factually, these effects are the basic
reason for the complex oscillation behavior. A common phenomenon in these
dynamical systems is that they seem to behave chaotically during some
transient periods, but eventually fall onto periodic stable attractors. This
has been called as the transient chaos or chaotic transient. Superlong
transient chaos occurs commonly in dissipative dynamical system, in this
case, oscillations starting from random initial conditions oscillate
chaotically for a very long time before they set into the final attractors
which are usually regular and stable [22,23]. In our system, we also find
the transient chaos and final attractors. Using the SPS of $(z,dz/dt)$, we
exhibit the attracting process of the transient chaos and the fixed points
of the final attractors. The phase trajectories of the final attractors are
also showed.

For a certain damping parameter $\eta $, and fixed value of parameter $%
\Lambda $, $\Delta E_{0}$, there exist many types of attractors when $\Delta
E_{1}$ is changed. For the same $\Delta E_{1}$, different initial conditions
will lead different final states. Starting from Rabi oscillation, with $%
z(0)=0.5$, $\phi (0)=0.0$, $\Delta E_{0}=0.0$, $\Lambda =10.0$, $\omega
=4\pi $ and $\eta =0.01$, for different parameter $\Delta E_{1}$, the SPS of
the attracting processes and the final attractors, and the phase
trajectories of the final attractors are presented in Fig. 8, the left
column shows the SPS of the attracting processes, the right column shows the
phase trajectories and SPS of the final states, (A) and (B) for $\Delta
E_{1}=3.000$, (C) and (D) for $\Delta E_{1}=7.500$. In the SPS, after the
transient chaos, the sampled points gradually come to the final fixed
points. The phase trajectories of final oscillations are closed curves, and
the corresponding SPS only contain fixed points which are noted as small
circles, so the final oscillations are frequency-locked (FL). When $\Delta
E_{1}=3.000$, there is only a single fixed point in the SPS, the
corresponding final oscillation is a period-one limit-cycle with frequency $%
\omega $; while for $\Delta E_{1}=7.500$, there exist five fixed points, and
then the final oscillation is a $\frac{1}{5}$ FL motion, this means the
oscillating frequency is $\frac{1}{5}\omega $. Fig. 9 presents the similar
dynamics starting form MQST, with $z(0)=0.75$, $\phi (0)=0.0$, $\Delta
E_{0}=0.0$, $\Lambda =10.0$, $\omega =2\pi $ and $\eta =0.001$, for
different parameter $\Delta E_{1}$. Where, (A) and (B) for $\Delta
E_{1}=1.000$, (C) and (D) for $\Delta E_{1}=1.700$. The transient chaos and
the FL oscillations appear too. When $\Delta E_{1}=1.000$, the eventual
oscillation is a period-one limit cycle with a non-zero time-averaged value
of the fractional population imbalance $z$, so the atoms are localized on
one of the condensates. Amazedly, for large $\Delta E_{1}$ (1.700), due to
the damping effects, the final $\frac{1}{6}$ FL oscillation possesses of a
non-zero time-averaged value of $z$, comparing with the non-damping regime
(Fig. 6), one can obtain that the proper damping can keep the MQST
long-lived.

\section{Summary and discussion}

Using both analytical and numerical methods, we analyzed the chaotic
oscillations between two coupled Bose-Einstein condensates with
time-dependent asymmetric trap potential. The trap asymmetry has been chosen
as $\Delta E=\Delta E_{0}+\Delta E_{1}\sin \omega t$, this can be realized
by varying the laser barrier position of the laser beam which possesses of
fixed value of intensity. The damping of the oscillations of the fractional
population imbalance is in form of $-\eta z(t)$, it commonly exists in the
two interacting condensates with different hyperfine levels in a single
harmonic trap.

In the perturbative regime, the population oscillations have been depicted
with Duffing equation, the chaotic oscillations near the separatrix solution
are detailed. The form of the general solution and the sufficient-necessary
conditions for stable oscillations are obtained. These conditions
sensitively depend on the initial conditions and the physical parameters,
and the first necessary condition indicates that the Melnikov function of
system is equal to zero, so the stable oscillations are Melnikov chaotic.
The stable curves are presented out for different initial conditions, the
sensitive dependence exhibits implicitly in the figures. Varying the damping
strength, the regions of chaotic oscillations and regular oscillations can
be changed into the other.

However, the general oscillations are not close to the separatrix solution,
and the usual parameters are not in perturbative regime, in such case, the
numerical method is very useful. Using the fourth Runge-Kutta method with
variable step-width, the chaotic population oscillations are simulated by
straightforward numerical integration of the dimensionless motion equations.
When the damping disappears, with the increasing of $\Delta E_{1}$, the
regular oscillations gradually become chaotic, and in the completely chaotic
regime, the long-lived localization or MQST disappears. In the SPS, the
single regular island is separated into many little islands, and then all
islands are submerged into the chaotic sea. In the completely chaotic
oscillations, the long-term localization or MQST disappears and the
short-term localization or MQST can be changed from one of the BECs to the
other through the tunnel of the Rabi oscillation. When the damping exists,
due to the damping effects, the system is not a Hamiltonian system but a
dissipative one, and the volume of the phase space is reduced by time
evolution. Then the stationary chaos disappears, the transient chaos is a
common phenomenon before regular stable frequency locked oscillations.
Surprisingly, the proper damping strength can keep the localization or MQST
long-lived.

In experiments, the long-term average lifetime of the transient chaotic
oscillation requires the measurements must be long-term too. So the
prediction of the relation between the average lifetime of the transient and
the physical parameters $(\eta $, $\Delta E_{0}$, $\Delta E_{1}$ and $%
\Lambda )$ may be a practical problem. And if one want to observe the
long-lived localization or MQST, the understanding of the basins of
attraction of the eventual FL oscillations in the parameter space will give
some useful indication of how to choose the physical parameters. We will
report these results in other papers. 
\begin{figure}[tbp]
\caption{The changing of the shape of the effective potential $V$, (A) with
fixed $\Lambda =2.0$ and different values of $H$, (B) with fixed $H=0.5$ and
different values of $\Lambda $.}
\label{potential}
\end{figure}
\begin{figure}[tbph]
\caption{The stability curves for different initial conditions with $H=0.5$, 
$\Lambda =0.5$, $\Delta E_{0}=0.0$ and $\eta =0.5$.}
\label{stability curves}
\end{figure}
\begin{figure}[tbp]
\caption{The regions of chaotic oscillations for different values of the
damping parameter $\eta $, with $\Lambda =4.0$, $\Delta E_{0}=0$ and $H=0.5$%
. }
\label{regions}
\end{figure}
\begin{figure}[tbp]
\caption{The time evolution of the fractional population imbalance $z$ with $%
\Lambda =10$, the left column with initial conditions $z(0)=0.5$ and $\phi
(0)=0.0$, the right column with $z(0)=0.8$, $\phi (0)=0.0$. (A) and (E) with 
$\Delta E=0.0$ and $\eta =0.0$; (B) and (F) with $\Delta E=0.0$ and $\eta
=0.5$; (C) and (G) with $\Delta E=1.0$ and $\eta =0.0$; (D) and (H) with $%
\Delta E=1.0$ and $\eta =0.5$.}
\end{figure}
\begin{figure}[tbp]
\caption{The stroboscopic Poincare section (SPS) of $(z,dz/dt)$ with $%
z(0)=0.5$, $\phi (0)=0.0$, $\Delta E_{0}=0.0$, $\Lambda =10.0$, $\omega
=4\pi $, $\eta =0.0$ and different values of $\Delta E_{1}$. }
\label{pssrb}
\end{figure}
\begin{figure}[tbp]
\caption{The stroboscopic Poincare section (SPS) of $(z,dz/dt)$ with $%
z(0)=0.75$, $\phi (0)=0.0$, $\Delta E_{0}=0.0$, $\Lambda =10.0$, $\omega
=2\pi $, $\eta =0.0$ and different values of $\Delta E_{1}$.}
\label{pssmq}
\end{figure}
\begin{figure}[tbp]
\caption{The completely chaotic oscillations and the corresponding power
spectra. The left column corresponds to $z(0)=0.5$, $\phi (0)=0.0$, $\Delta
E_{0}=0.0$, $\Lambda =10.0$, $\omega =4\pi $, $\Delta E_{1}=7.500$ and $\eta
=0.0$. The right column corresponds to $z(0)=0.75$, $\phi (0)=0.0$, $\Delta
E_{0}=0.0$, $\Lambda =10.0$, $\omega =2\pi $, $\Delta E_{1}=1.700$ and $\eta
=0.0$.}
\label{chaotic oscillation}
\end{figure}
\begin{figure}[tbp]
\caption{The stroboscopic Poincare section (SPS) of $(z,dz/dt)$ and
frequency-locked oscillations with $z(0)=0.5$, $\phi (0)=0.0$, $\Delta
E_{0}=0.0$, $\Lambda =10.0$, $\omega =4\pi $, $\eta =0.01$ and different
values of $\Delta E_{1}$. (A) and (B) with $\Delta E_{1}=3.000$, (C) and (D)
with $\Delta E_{1}=7.500$.}
\label{drbfl}
\end{figure}
\begin{figure}[tbp]
\caption{The stroboscopic Poincare section (SPS) of $(z,dz/dt)$ and
frequency-locked oscillations with $z(0)=0.75$, $\phi (0)=0.0$, $\Delta
E_{0}=0.0$, $\Lambda =10.0$, $\omega =2\pi $, $\eta =0.001$ and different
values of $\Delta E_{1}$. (A) and (B) with $\Delta E_{1}=1.000$, (C) and (D)
with $\Delta E_{1}=1.700$.}
\label{dmqfl}
\end{figure}

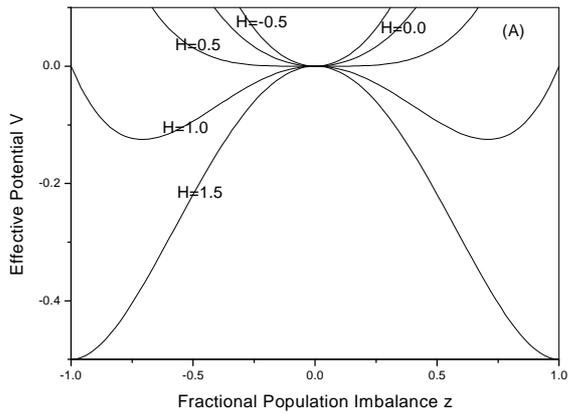
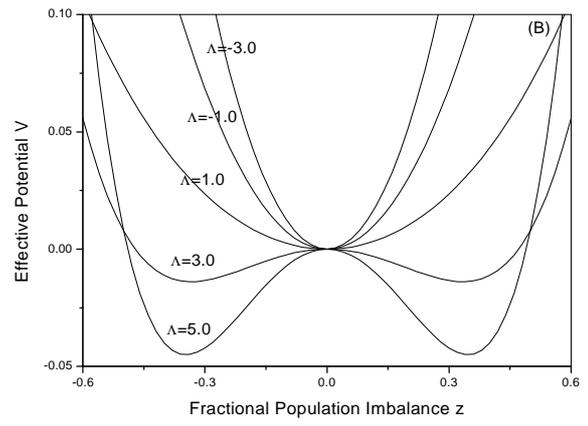
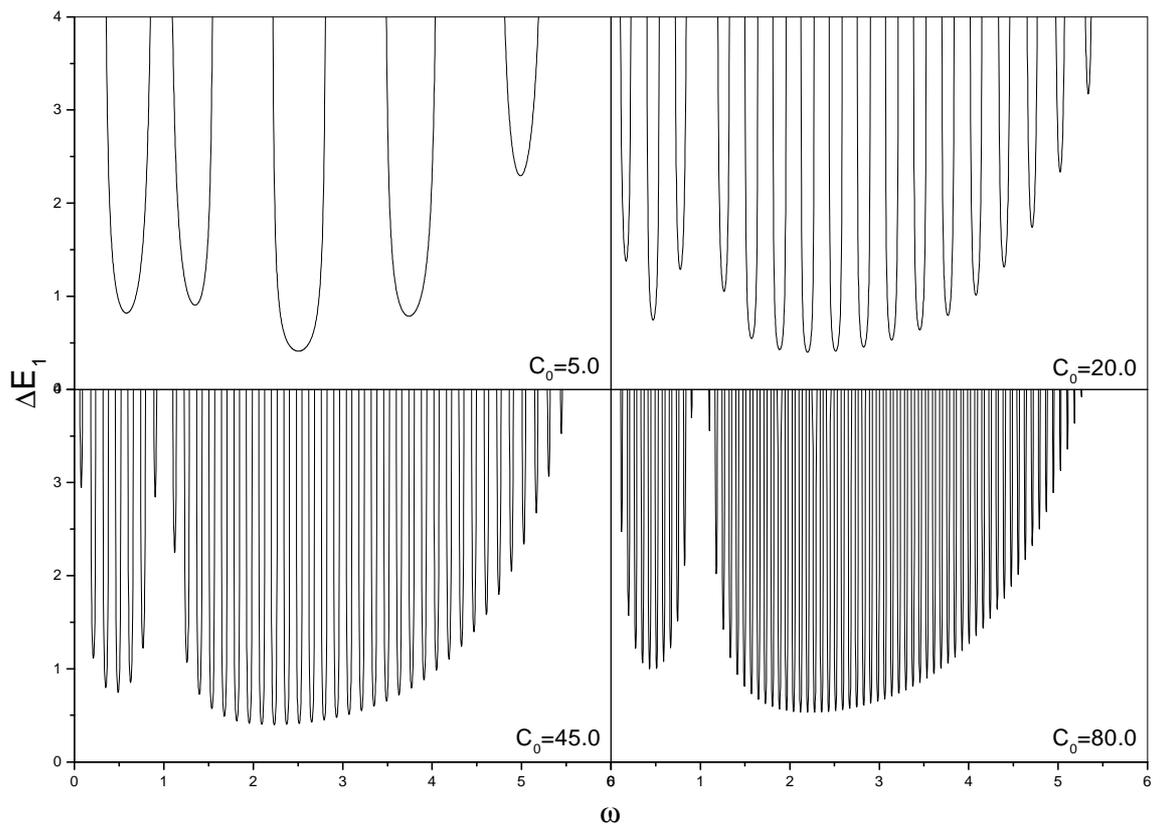
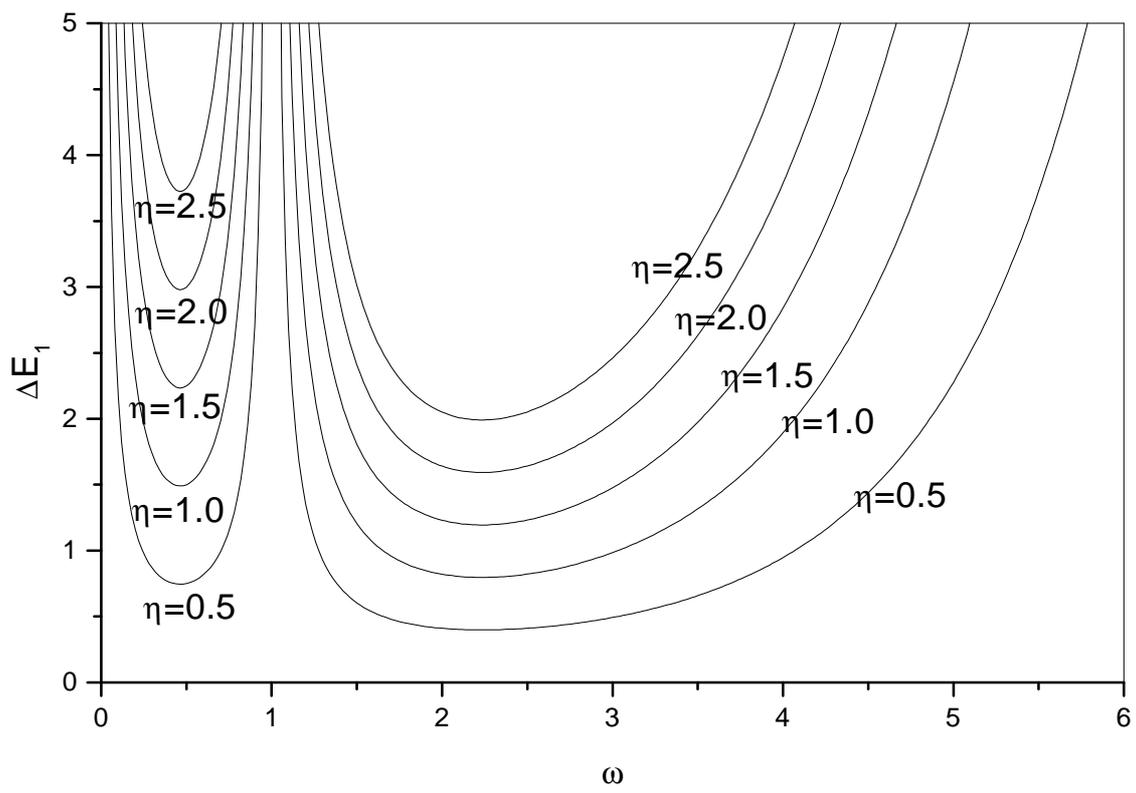
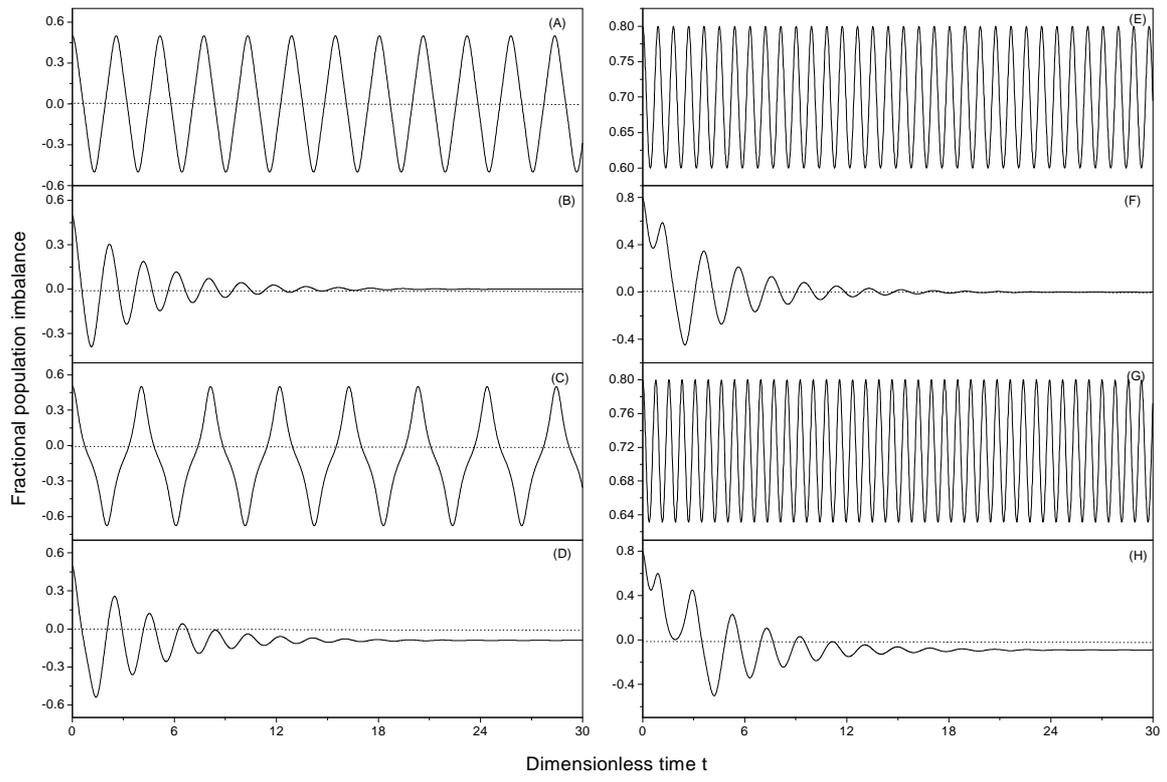
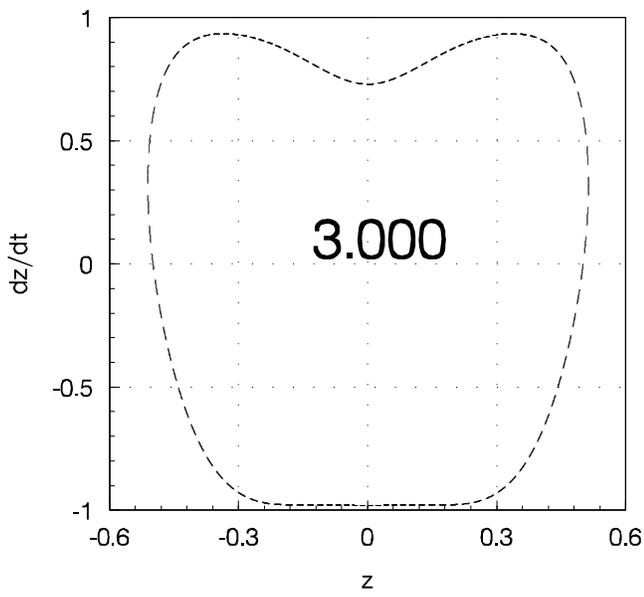
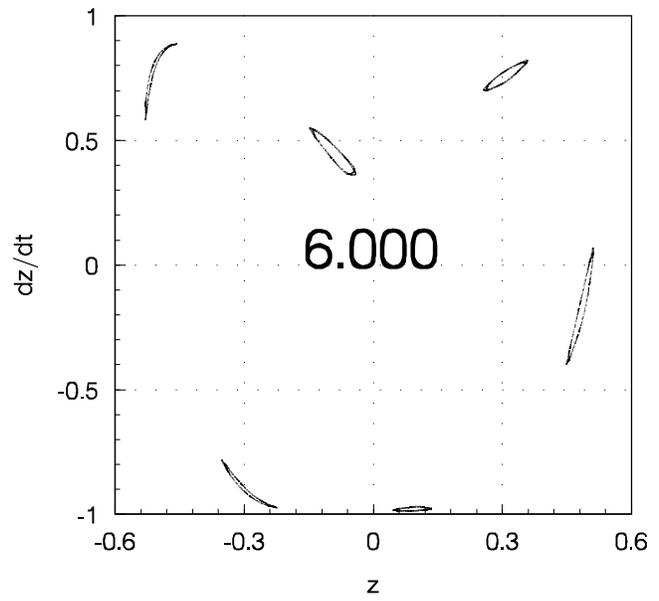
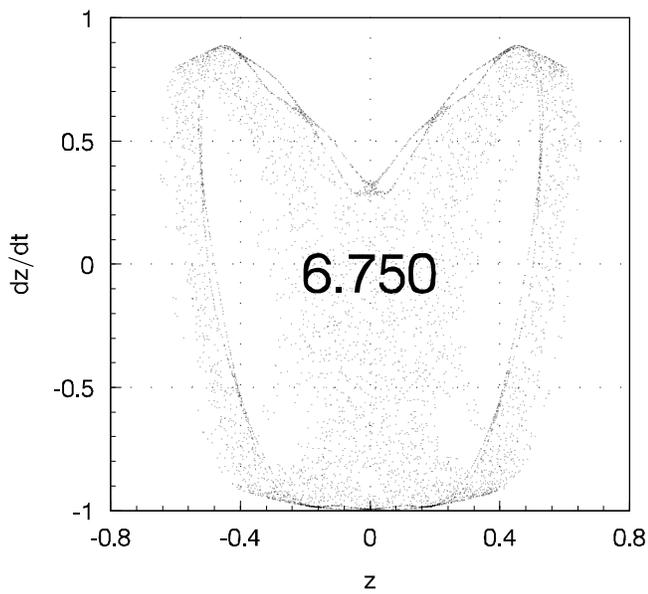
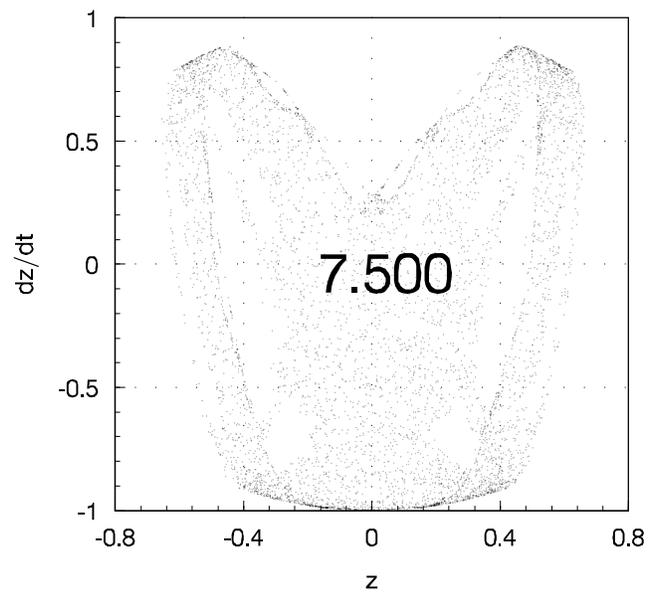
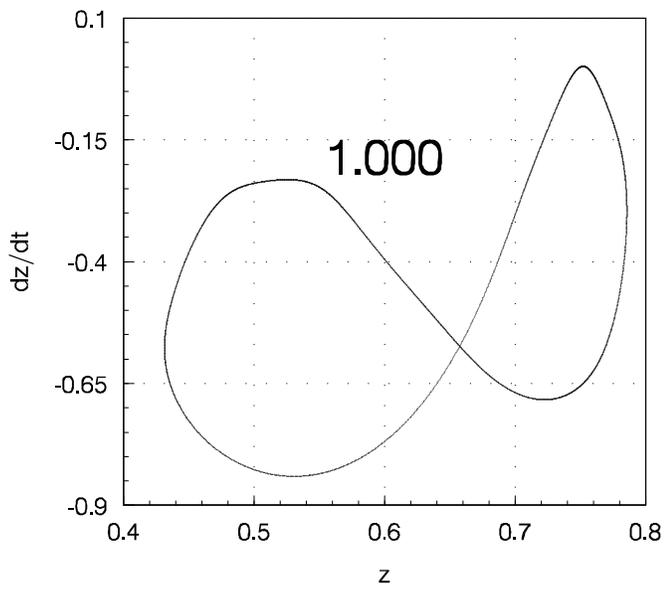
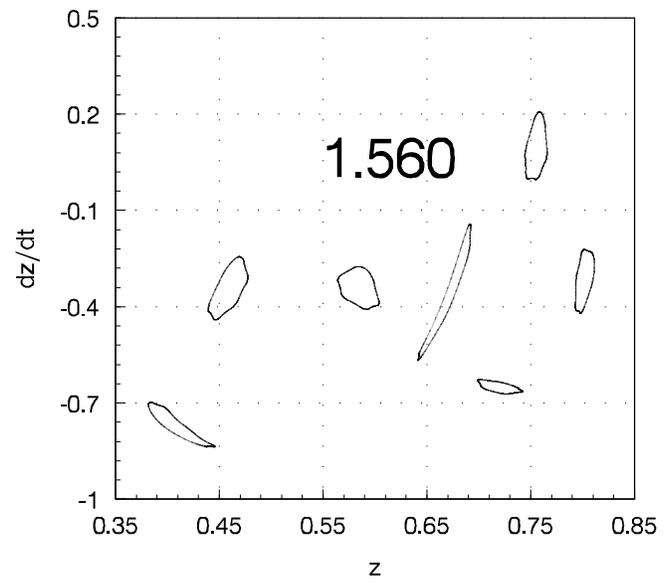
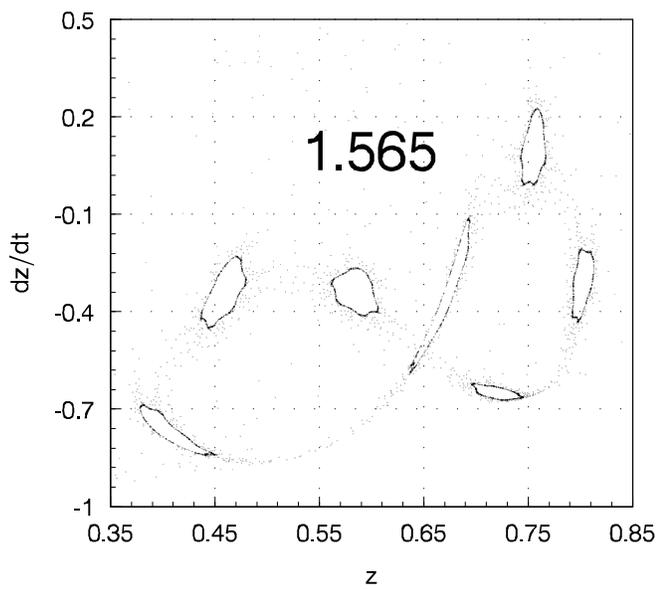
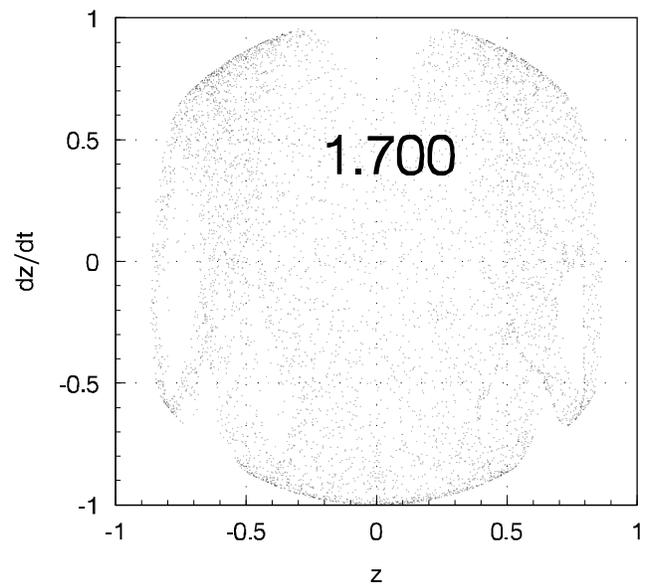
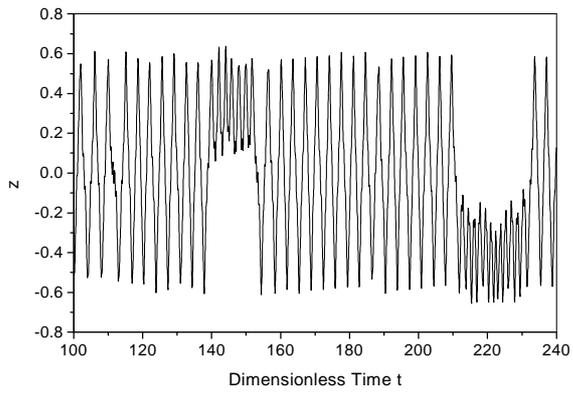
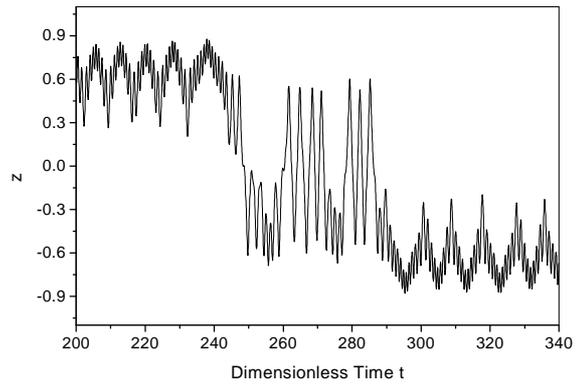
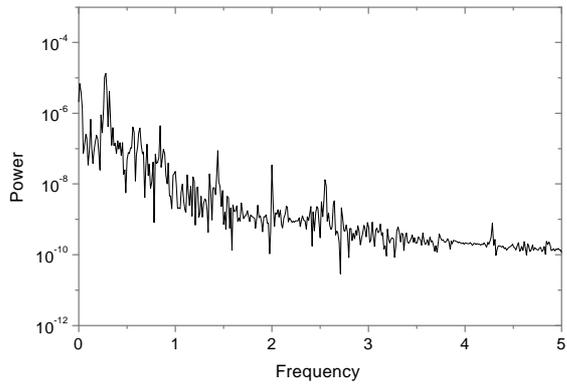
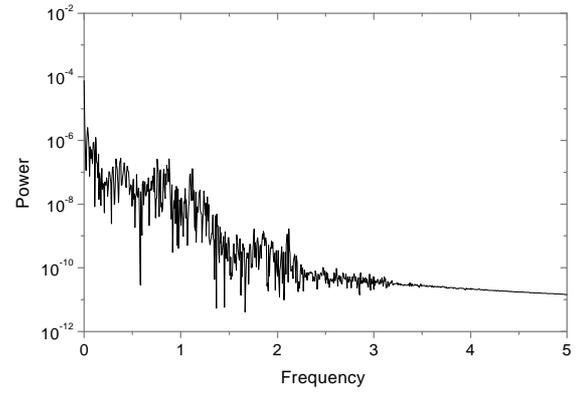
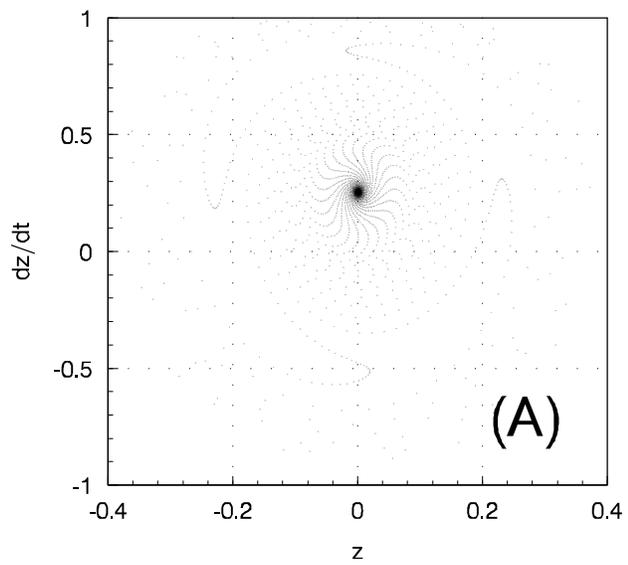
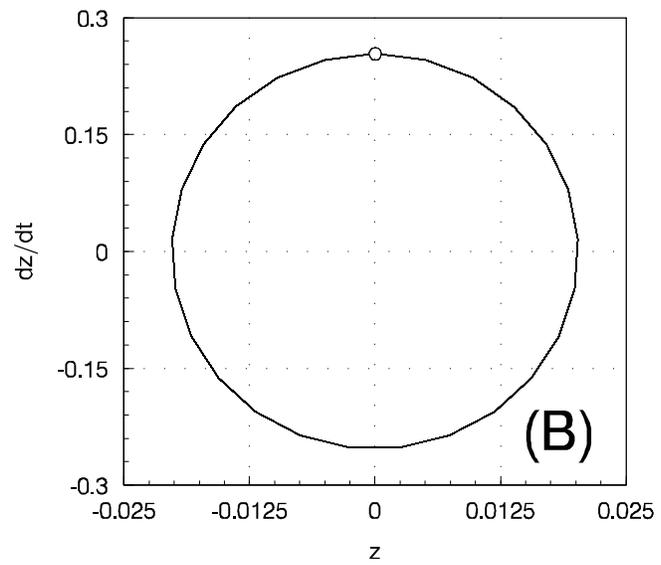
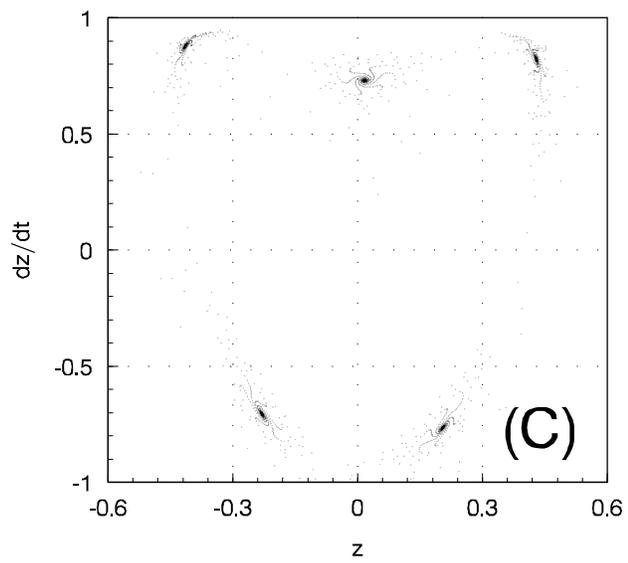
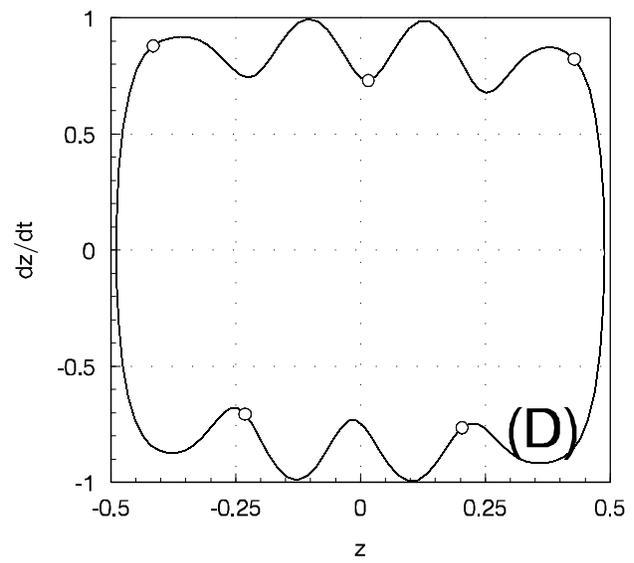
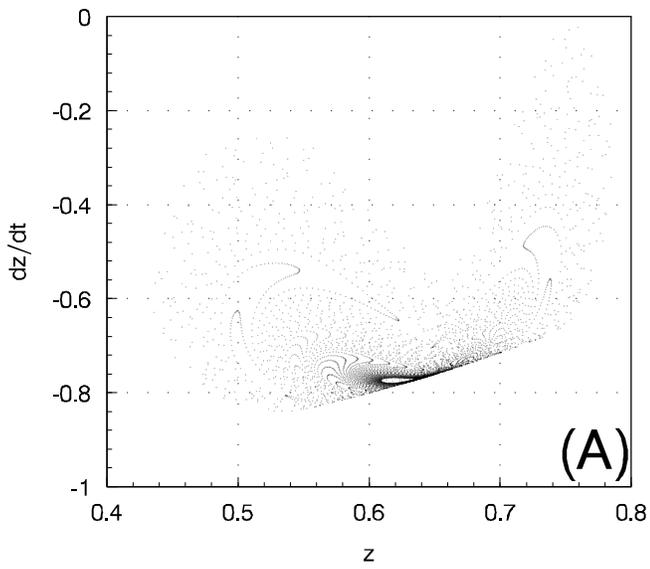
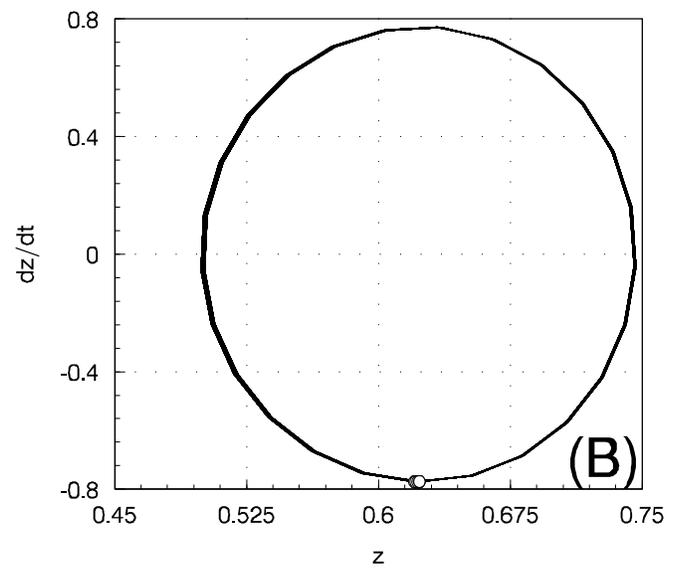
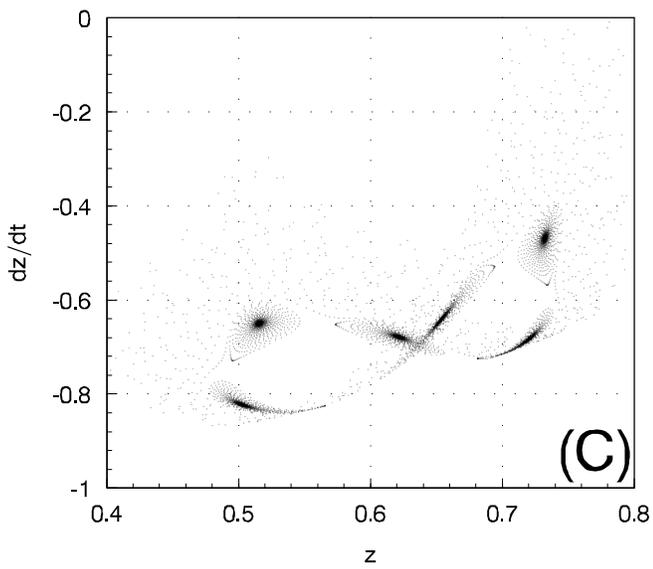
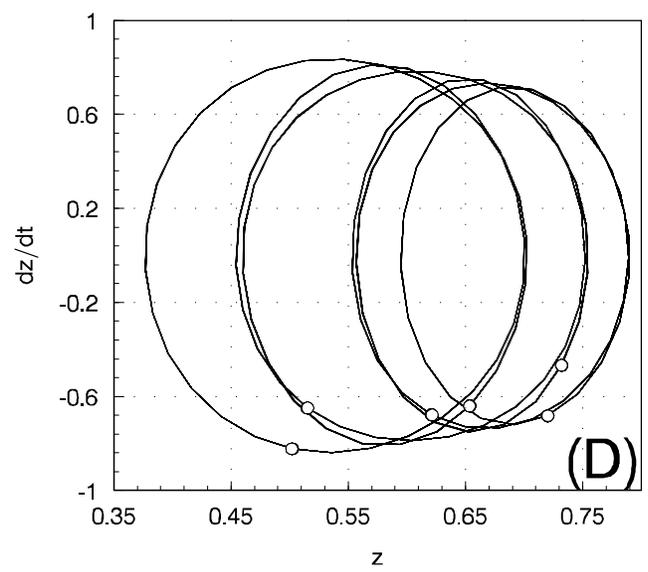
\end{document}